\documentclass[twocolumn,prb]{revtex4}

  \usepackage[dvips]{graphicx}

\usepackage{dcolumn}
\usepackage{amsmath}
\usepackage{latexsym}
\usepackage{color}
\usepackage{ulem}

\usepackage{footmisc}

\begin{document}

\title[]{Quasi-phasematching in a poled Josephson traveling-wave parametric
amplifier with three-wave mixing}
\author{A. B. Zorin}
\email{alexander.zorin@ptb.de}
\noaffiliation
\affiliation{Physikalisch-Technische Bundesanstalt, Bundesallee
100, 38116 Braunschweig, Germany}
\noaffiliation

\date{May 15, 2021}

\begin{abstract}
We develop the concept of quasi-phasematching (QPM) by implementing
it in the recently
proposed Josephson traveling-wave parametric amplifier (JTWPA)
with  three-wave mixing (3WM). The amplifier is
based on a ladder transmission line consisting of flux-biased
radio-frequency SQUIDs whose nonlinearity is of $\chi^{(2)}$-type.
QPM  is achieved in the 3WM process,
$\omega_p=\omega_s+\omega_i$ (where $\omega_p$, $\omega_s$, and $\omega_i$
are the pump, signal, and idler frequencies, respectively)
due to designing the JTWPA to include periodically inverted
groups of these SQUIDs that reverse the
sign of the nonlinearity. Modeling shows
that the JTWPA bandwidth is relatively large (ca. $0.4\omega_p$)
and flat, while unwanted modes, including
$\omega_{2p}=2\omega_p$, $\omega_+=\omega_p +\omega_s$,
$\omega_- = 2\omega_p - \omega_s$, etc.,
are strongly suppressed with the help of engineered dispersion.


\end{abstract}

\maketitle



Due to vanishingly small losses and an ultimately quantum level
of internal noise, \cite{Louisell1961}
cryogenic traveling-microwave parametric amplifiers based on the kinetic
inductance of superconducting wires \cite{Eom2012,Vissers2016,Malnou2021} and
Josephson junctions \cite{Macklin2015,White2015,Planat2020,Ranadive2021} are
considered highly useful quantum devices that can be applied in
precision quantum measurements, photon detection,
quantum communication and quantum computing. \cite{Wallraff2004,Devoret2013}
In comparison to their cavity-based counterparts,
\cite{Movshovich1990,Castellanos-Beltran2007,Yamamoto2008}
traveling-wave parametric amplifiers have the advantages of a wider
bandwidth and a larger dynamic range. However, realization
of phase matching that is
sufficient for ensuring large (ideally, exponential \cite{Agrawal2007})
flat signal gains in a wide bandwidth remains the greatest challenge when designing such
amplifiers. \cite{Macklin2015,White2015,Planat2020,OBrien2014,Bell-Samolov2015,WenyuanZhang2017,Ranadive2021}

\makeatletter{\renewcommand*{\@makefnmark}{}
\footnotetext{This article may be downloaded for personal use only.
Any other use requires prior permission of the author and AIP Publishing.
This article appeared in Appl. Phys. Lett. \textbf{118}, 222601 (2021)
and may be found at https://doi.org/10.1063/5.0050787
}\makeatother}

Josephson traveling-wave parametric
amplifiers (JTWPAs) have the architecture
of a transmission line with discrete elements. \cite{Mohebbi2009}
Their performance is normally based on a Kerr-like nonlinearity, i.e.,
a Josephson relation between a small current $I$ and a small flux $\Phi$
of the form $I = (\Phi - \gamma \varphi_0^{-2} \Phi^3)L^{-1}_0$, where $L^{-1}_0$ is the
inverse inductance, $\varphi_0 = \Phi_0/2\pi = \hbar/2e$
is normalized flux quantum, and $\gamma$ $(> 0)$ is the
Kerr coefficient. \cite{Yaakobi2013}
These amplifiers are operated in a four-wave mixing (4WM)
regime wherein the pump ($\omega_p$),
signal ($\omega_s$), and idler ($\omega_i$) frequencies obey the relation
$ 2\omega_p - \omega_s -\omega_i =  0$.
However, here, it is difficult to fulfill
the phase-matching relation
for corresponding wavenumbers, $\Delta k =2 k_p - k_s - k_i = 0$, in a wide range of
frequencies and pump powers. Specifically, the Kerr effect
causes unwanted self-phase modulation (SPM) and cross-phase modulation (XPM)
of waves because of the
intensity-dependent phase velocities leading
to phase shifts. \cite{Agrawal2007}
This leads to imperfect phase matching, $\Delta k \neq 0$,
which can only be improved by means of dispersion engineering.
This can be done, for example, by applying resonant phase matching
\cite{White2015,Macklin2015,OBrien2014}
or by inverting the Kerr coefficient sign, $\gamma < 0$. \cite{Bell-Samolov2015,Ranadive2021}

Recently, superconducting elements that possess widely tunable
non-centrosymmetric nonlinearity  of the $\chi^{(2)}$-type
have been proposed.
Such elements (see Fig.~1) include either non-hysteretic
radio-frequency SQUIDs (rf-SQUIDs)
\cite{Zorin2016} or asymmetric multijunction SQUIDs, i.e.,
so-called superconducting nonlinear asymmetric inductive elements
(SNAILs). \cite{Frattini2017,Zorin2017}
An appropriate magnetic flux $\Phi_e$ applied to these SQUIDs allows
the current-flux relation to be adjusted to fit the form
\begin{equation}
I = (\Phi - \beta \varphi_0^{-1} \Phi^2)L^{-1}_0,
\label{beta-nonlinearity}
\end{equation}
where coefficient $\beta$ is the electrical analog of the susceptibility tensor
$\chi^{(2)}$ in the optics. \cite{Agrawal2007}
Formula~(\ref{beta-nonlinearity}) provides the means for
designing Kerr-free ($\gamma = 0$) JTWPAs with pure three-wave
mixing (3WM), 
$\omega_p - \omega_s -\omega_i = 0$. \cite{Zorin2016,Zorin2017,Miano2018}
Here, the pump frequency $\omega_p$ favorably lies
outside the signal band, $(0,~ \omega_p)$.

Another remarkable property of 3WM is the absence of SPM and XPM effects.
Hence, for a sufficiently
small dispersion, the
phase-matching condition, $k_p - k_s - k_i \approx 0$, can be met
in a wide frequency range. The dispersion relation in the
ladder-type transmission line is\:\cite{Zorin2019}
\begin{equation}
k = 2 \arcsin \left( \frac{\omega/2\omega_0}{\sqrt{1-\omega^2/\omega_J^2}} \right)
\approx  \frac{\omega}{\omega_0} \left(1+ \frac{\omega^2}{2\omega_J^2}
+ \frac{\omega^2}{24\omega_0^2} \right),
\label{dispersion}
\end{equation}
where $k$ is the wavenumber normalized on the reverse size
of elementary cell $d^{-1}$ and the frequencies $\omega_J = 1/\sqrt{LC_J}$ and
$\omega_0 = 1/\sqrt{LC_0}$ (see
notations in Fig.\,1) are the
SQUID plasma frequency and the line cutoff frequency, respectively. Thus,
the condition of small dispersion reads $\omega_p \ll \omega_J, \omega_0$.

\begin{figure}[t]
\begin{center}
\includegraphics[width=3.5in]{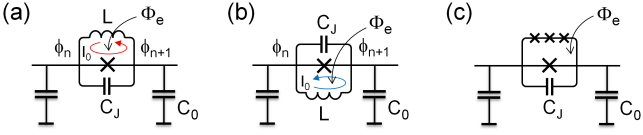}
\caption{Fragments of superconducting transmission lines
with nonlinearities of the $\chi^{(2)}$-type due to flux-biased rf-SQUIDs
(Josephson junctions are denoted by crosses).
Panels (a) and (b) show inverted configurations of rf-SQUIDs, which have opposite signs
of coefficient $\beta$ in a uniform external magnetic field.
In both cases, (a) and (b), the circulating current $I_0$ is  anticlockwise.
The screening parameter value in these rf-SQUIDs, $\beta_L = L I_c/\varphi_0 < 1$,
where $I_c$ is the Josephson critical current.\cite{Zorin2016} (c) SNAIL circuit in which a
short array of Josephson junctions that forms the upper SQUID branch
serves as the equivalent (kinetic) inductor $L$. \cite{Frattini2017,Zorin2017}
}
\label{poling-concept0}
\end{center}
\end{figure}

In general, mixing of the signal and the pump
using the nonlinearity given by Eq.~(\ref{beta-nonlinearity})
results not only in the idler wave being generated with difference frequency
$\omega_i =  \omega_p - \omega_s < \omega_p$ but also
in waves with frequencies above $\omega_p$,
including $\omega_+ = \omega_p + \omega_s$,
$\omega_-  = 2\omega_p - \omega_s = \omega_p + \omega_i$, $2\omega_p$.
Moreover, for vanishingly small dispersion, phase-matching conditions
for these accompanying processes may also be met.
In the original concept of the parametric amplifier
with 3WM, \cite{Tien1958}$^,$\cite{Cullen1960}
these waves were excluded, so their effects on the basic waves
were not considered. However, later analysis
showed \cite{Erickson2017,Zorin2019,Dixon2020} that these modes not only
absorb pump energy (in accordance with the Manley-Rowe
relations \cite{Manley-Rowe1956}), but also
interact with signal and idler waves, causing notable signal-rise undulation.
\cite{Zorin2019} As a result, premature pump depletion and irregular dependence
of the gain on the signal frequency and pump power were observed in both
the simulations \cite{Dixon2020} and the experiment. \cite{Zorin2017,Miano2018}

A straightforward method of suppressing the propagation of unwanted high-frequency
modes is to reduce the cutoff frequency $\omega_0$ in such a way that
$\omega_p \lesssim \omega_0 < \omega_\pm,\, \omega_{2p}$.
However, this approach has several drawbacks: First, because the
wavelength $\lambda = 2\pi \omega_0/\omega$ is only about the size of 10 cells,
the discreteness of the line becomes critical;
thus, its transmission and, hence, the gain
may exhibit notable ripples. Second, in order to keep the impedance
unchanged, i.e., $Z_0 = \sqrt{L/C_0} = 50~\Omega$, both the inductance $L$ and
the ground capacitance $C_0$
should be increased;  this is possible only by increasing their physical
sizes, which may entail difficulties in fabricating the circuit on a standard chip.
Third, the maximum pump power is
$P_p^{\textrm{max}} \approx (\omega_0 \varphi_0)^2/2Z_0$; \cite{Zorin2016}
thus, a small value of $\omega_0$ will yield a small $P_p^{\textrm{max}}$
and, hence, cause pump depletion at a rather low signal power.

\begin{figure}[b]
\begin{center}
\includegraphics[width=3.5in]{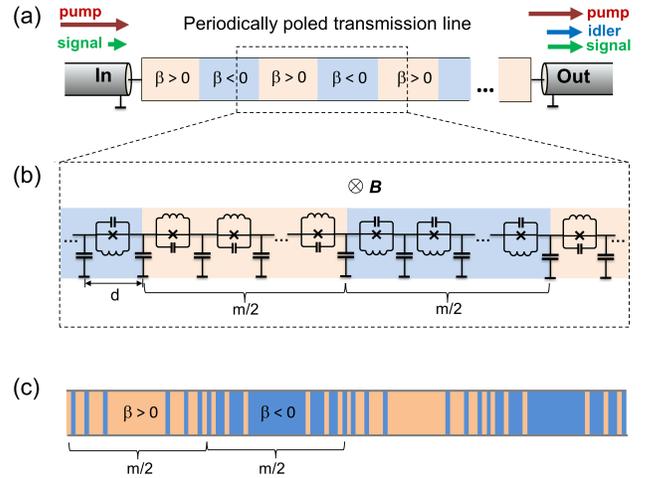}
\caption{(a) Schematic circuit diagram of a QPM-based JTWPA comprising a serial
array of $N~ (\sim 10^3)$ flux-biased rf-SQUIDs with a periodically inverted sign
of nonlinear coefficient $\beta$.
(b) Poled segments of the line that have different signs
of coefficient $\beta = |\beta| \textrm{sgn} [\sin(2 \pi n/m)]$, which are achieved
due to the inversion of rf-SQUIDs in a uniform magnetic field $\textit{\textbf{B}}$.
(c) A pulse-width modulated profile that synthesizes a smooth
variation of the effective local value of coefficient $\beta(x)$;
the different segments, which have opposite signs of $\beta$, are shown by means of
different colors.}
\label{poling-concept}
\end{center}
\end{figure}

To solve this problem,
we first engineered a reasonably large dispersion.
By setting the SQUID plasma frequency slightly above the double
pump frequency, $\omega_J \gtrsim 2\omega_p$, which is done by
increasing the SQUID capacitance $C_J$, a phase mismatch for unwanted
modes, $\omega_+$, $\omega_-$, and $\omega_{2p}$  (all of which are between $\omega_p$
and $2\omega_p$) can be made sufficiently large in accordance with formula (\ref{dispersion}).
The unavoidable phase mismatch for basic 3WM, $\Delta k_0
= k_p - k_s-k_i > 0$, is not dramatically large in this case,
although it depends on the signal frequency $\omega_s$. \cite{Zorin2019}
For example, for $\omega_s\approx\omega_i\approx0.5\omega_p$, the mismatch is at its
maximum, $\Delta k_{0} \approx 3\omega_p^3/8\omega_J^2\omega_0
\sim 3\omega_p/32\omega_0$. This value corresponds to a reasonably
large coherence length of $\ell_c=\pi/\Delta k_0 \sim 10^2$ cells.
In principle, this mismatch can be improved by applying either resonator
phase-matching \cite{OBrien2014,White2015,Macklin2015,Dolata2020}
or periodic loading of the line. \cite{Eom2012,Erickson2017,Malnou2021}

In this Letter, however, we use so-called quasi-phasematching
(QPM), \cite{Armstrong1962,Boyd2008}
whose possible implementation in our JTWPA is outlined in Fig.~2.
QPM was originally exploited in optical devices, including parametric amplifiers,
\cite{Charbonneau-Lefort2008} parametric oscillators, \cite{Byer1997}
second-harmonic generators, \cite{Fejer1992,Yamada1993} and spontaneous parametric
down-converters. \cite{Fiorentino2007} These devices were based on $\chi^{(2)}$ nonlinearity
in a poled optical material like LiNbO$_3$.

The core operating principles of QPM are as follows.
After passing coherence length $\ell_c$ each time the signal wave is about to begin to decrease
(as a consequence of the wavenumber mismatch), a reversal of the $\chi^{(2)}$-sign occurs,
which changes the signal phase by $\pi$ and allows the wave amplitude to continue to grow.
When another $\pi$ phase shift has accumulated, the constant-sign domain is reversed
again, and so on.
Thus, due to the periodic inversion of the orientation (poling) of the LiNbO$_3$ crystal,
an effective compensation of the phase
mismatch $\Delta k_0$ in 3WM is possible. \cite{Boyd2008}

The mechanism of QPM can be illustrated using simplified
coupled-mode equations (CMEs) \cite{Agrawal2007}
for slowly varying complex
amplitudes of the signal and idler waves, $A_s$ and  $A_i$, respectively,
which are coupled to the pump wave with amplitude $A_p$.
The corresponding signal (idler) wave is given by the expression
$\phi_{s,i}(x,t) = 0.5 A_{s,i}(x) e^{i(k_{s,i}x - \omega_{s,i}t)} + \textrm{c.c.}$,
where $x$ is a continuous coordinate that is normalized on cell size $d$.
Thus, the superconductor phase $\phi_{s,i}(x,t)$
on the $n$-th node at $x=n$ takes the value
$\phi_n^{(s,i)}(t) = \Phi_n^{(s,i)}(t)/\varphi_0$. \cite{Zorin2016}
Treating the pump as undepleted, $|A_p(x)| = A_{p0} \gg |A_{s,i}(x)|$, the
CMEs for Kerr-free ($\gamma = 0$) 3WM read
\begin{eqnarray}
\frac{d A_{s}}{dx} = -0.5 \eta(x) |\beta| k_p k_{i} A_{p0}
 A^*_{i}(x) e^{i \Delta k_{0}x}, \label{CME-3WM-1} \\
\frac{d A_{i}}{dx} = -0.5 \eta(x) |\beta| k_p k_{s} A_{p0}
 A^*_{s}(x) e^{i \Delta k_{0}x}.
\label{CME-3WM-2}
\end{eqnarray}
The peculiarity of these otherwise standard \cite{Agrawal2007}
equations is the presence of modulation term $\eta(x)$,
a periodic function with a period of $x_m =m$. This function takes
only binary values, i.e., 1 or $-1$, and thus determines the sign of $\beta$.
To capture the QPM effect, we approximate
the function $\eta(x)$ by the
dominant (first) term in its Fourier series expansion, \cite{Byer1997} i.e.,
$\eta(x) \approx \eta_0 \sin k_mx$.
Then, the exponent entered on the right-hand side
of Eqs.~(\ref{CME-3WM-1},\ref{CME-3WM-2}), takes the form
\begin{equation}
\eta(x) e^{i \Delta k_{0}x} = \frac{\eta_0}{2i}
[e^{i (\Delta k_{0}+k_m)x} - e^{i (\Delta k_{0}-k_m)x}].
\label{exp}
\end{equation}
Choosing a poling period $m$ in such a way
that $k_m \equiv 2\pi/m= \Delta k_{0}$, i.e., the actual QPM condition,
the first (fast oscillating) term on the right-hand
side of Eq.~(\ref{exp}) can be
omitted as its integral effect is apparently small,
while the second term, $e^{i (\Delta k_{0}-k_m)x} =1$,
enables efficient parametric mixing.
As a result, an exponentially increasing solution, $|A_{s,i}|\propto e^{gx}$, in which the
exponential gain factor is
$g = 0.25 k_p \sqrt{k_s k_i} \beta_{\textrm{eff}} A_{p0}$, takes place
on a much larger scale than $m = 2\pi/k_m = 2\ell_c$. \cite{Boyd2008}
The effective value of the nonlinear coefficient
is $\beta_{\textrm{eff}} = 0.5 \eta_0 |\beta|$.
If the function $\eta(x)$ has a meander shape with a 50\% duty cycle,
the Fourier coefficient is $\eta_0 = 4/\pi$, yielding
$\beta_{\textrm{eff}} = (2/\pi) |\beta| < |\beta|$.  \cite{Boyd2008}
The resulting reduction in gain can be compensated by applying a somewhat
larger pump amplitude $A_{p0}$ and/or an increasing length $N$.

\begin{figure}[b]
\begin{center}
\includegraphics[width=3.4in]{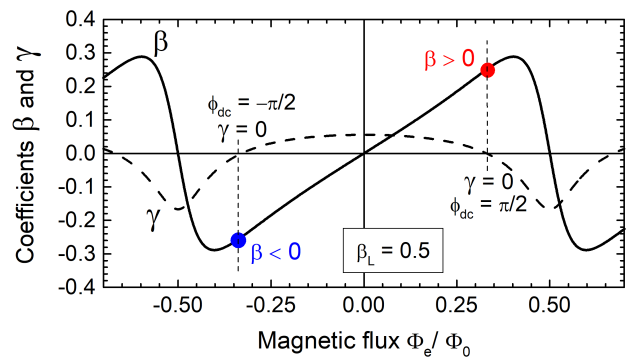}
\caption{Nonlinear coefficients $\beta$
(solid curve) and $\gamma$
(dashed curve) versus external magnetic flux $\Phi_e$.
The red and blue solid circles indicate the optimal working points which correspond to
$\phi_{\textrm{dc}} = \pm \pi/2$ (yielding
zero Kerr coefficient, $\gamma = 0$)
for rf-SQUID configurations shown in Figs.~1(a)
and 1(b), respectively. Note that the inductance of the Josephson junction
at these points is infinite, while the total inductance of the rf-SQUID
is given by the linear inductance $L$,
$L_0 = L/(1+\beta_L \cos \phi_{\textrm{dc}})= L$. \cite{Zorin2017}}
\label{beta-gamma}
\end{center}
\end{figure}

In contrast to the optical-material QPM technologies,
\cite{Armstrong1962,Byer1997,Fiorentino2007}
the rf-SQUID-based transmission line can be  poled in a relatively
simple way. A perpendicular magnetic field $\textbf{\textsl{B}}$
creates a magnetic flux $\Phi_e=|\textbf{\textsl{B}}|S$ that is applied to each
SQUID; here, $S$ is the loop area.
Assuming without loss of generality that $0<\Phi_e \leq 0.5 \Phi_0$, one
obtains an anticlockwise direction for the circulating current $I_0$ independently of
the SQUID configuration (as shown in Figs.~1(a) and 1(b)).
Then, defining the constant phase drop on the Josephson junction as
$\phi_{\textrm{dc}} = \phi_n - \phi_{n+1}$, we arrive
at an rf-SQUID equation for the magnetic flux \cite{Likharev1986}
in the form
\begin{equation}
\Phi_e/\varphi_0 = \pm \phi_{\textrm{dc}} \pm \beta_L \sin\phi_{\textrm{dc}},
\label{fluxPhiE-phiDC}
\end{equation}
where the sign $+$ ($-$) corresponds to the configuration shown in Fig.~1(a) (1(b)).
A change in the sign of the right-hand side of this
equation is equivalent to a
change of $\Phi_e \rightarrow - \Phi_e$ (or, equivalently, a virtual flip
of the magnetic field $\textbf{\textsl{B}}$). As a result, the optimal phase drop,
$\phi_{\textrm{dc}} = \pi/2$, which ensures
both a zero Kerr coefficient $\gamma$
(even periodic function of flux $\Phi_e$)
and nearly the maximum of coefficient $\beta$ (odd periodic
function of $\Phi_e$), changes its value, $\pi/2 \rightarrow -\pi/2$
(see Fig.~\ref{beta-gamma}).
Thus, the sign of nonlinearity constant $\beta$ is
flipped, $\beta \rightarrow -\beta$.

To demonstrate the efficiency of the QPM concept for JTWPA
with 3WM, we modeled the
circuit with the architecture shown in Fig.~2, neglecting possible small losses.
We designed a sufficiently low plasma frequency $\omega_J$, i.e.,
$2\omega_p \lesssim \omega_J \ll \omega_0$.
For the analysis of this circuit,
we used the set of six CMEs \cite{Zorin2021}  (the so-called CME-2 set \cite{Dixon2020}),
which takes into consideration all significant
modes with frequencies
below the actual transmission threshold of $\omega_J$, i.e.,
$\omega_s,~\omega_i,~\omega_p,~ \omega_+,~ \omega_-,$
and $\omega_{2p}$, and thus describes all relevant mixing processes
involving large pump. The set of these equations
was numerically solved \cite{Zorin2021}
by means of the standard Runge-Kutta method. The modeling showed
the maximum cross-gain
for modes $\omega_+$ and $\omega_-$ 
was safely smaller than the direct signal gain. The pump power
converted into the second harmonic, $\omega_{2p}$, was much less than 1\%.

\begin{figure}[b]
\begin{center}
\includegraphics[width=3.4in]{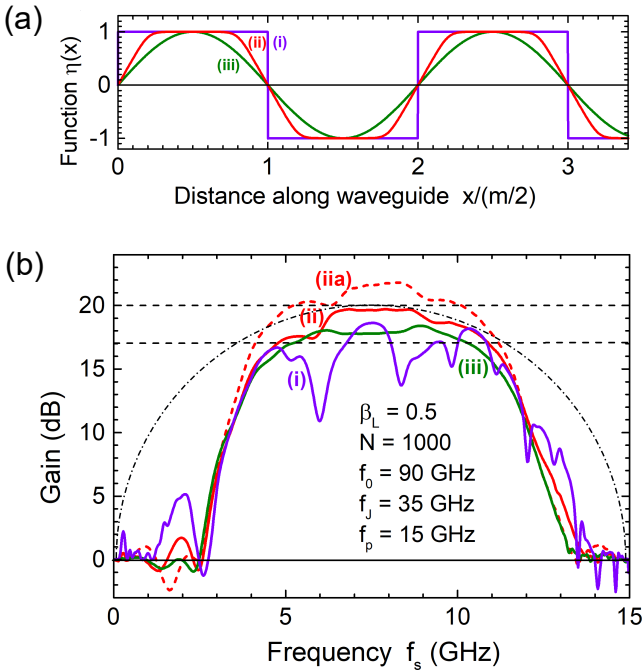}
\caption{(a) Three types of periodic poling shapes $\eta(x)/\eta_0$ that
enable QPM: (i) meander with a 50\% duty cycle (violet curve);
(ii) tapered meander (red curve); and (iii) sinusoidal shape (green curve).
(b) Gain versus signal frequency in quasi-phasematched JTWPAs
that have periodic poling of three types, (i)-(iii). The curves have
the same colors as in panel (a). The circuit parameters for the given
frequencies $\omega_0$, $\omega_J$, and $\omega_p$,  the
screening parameter $\beta_L$, and  the impedance $Z_0$ of $50\,\Omega$ are as
follows: $L = Z_0/\omega_0 \approx 88$\,pH, $C_0 = 1/\omega_0 Z_0 \approx 32$\,fF,
$C_J = 1/\omega_J^2 L \approx 235$\,fF, $I_c = \beta_L\varphi_0/L
\approx 1.8\,\mu$A, $|\beta| = 0.24$, $m = 448$ for solid curves (i), (ii), and (iii)
and $m = 420$ for dashed curve (iia). The undepleted pump power (ca. $-$70~dBm)
ensures that the
phase oscillations on every rf-SQUID have an amplitude of
$\phi_p = (\omega_p/\omega_0)A_{p0} \approx 0.5$ rad.
The dash-doted semiellipse shows the shape of the signal gain in an ideal 3WM
parametric amplifier that has perfect phase
matching in the full range $(0, \omega_p$) and thus
a wider bandwidth. \cite{Cullen1960}}
\label{poling-concept}
\end{center}
\end{figure}

Figure~4(a) shows three types of nonlinearity profiles
$\eta(x) = \beta(x)/\beta_{\textrm{max}}$ that have
similar periods $m$. This set includes
(i) a meander with a 50\% duty cycle, i.e.,
$\eta(x) = \textrm{sgn} [\sin(2\pi x/m)] $ (violet curve);
(ii) a tapered meander (widely used in optics for reducing
the gain ripple \cite{Charbonneau-Lefort2008}),
whose positive half-period $(0 \leq x \leq m/2)$
is described in our case by the formula
\begin{equation}
\eta(x) = - a \ln[ e^{-2x/ a b m}+ e^{-1/a} + e^{-(m-2x)/a b m}]
\label{smoothing}
\end{equation}
with fixed parameter values, $a=0.15$ and $b=0.2$ (red curve);
and (iii) a sinusoidal profile, $\eta(x) = \sin(2\pi x/m) $ (green curve).
To implement shapes (ii) and (iii), they were "digitized" using pulse-width
modulation, \cite{Zorin2021} resulting in a
variable density of $\beta$-flips per unit length. \cite{Huang2006}
The transmission line poled in this fashion is schematically
shown in Fig.~2(c).

The signal gain in JTWPA with $N = 1000$ and $\beta_L =0.5$,
which is calculated for poling profiles (i)-(iii),
is shown in Fig.\,\ref{poling-concept}(b) as a function
of frequency. The pump frequency and the line characteristic frequencies
were chosen in such a way
that $\omega_p/\omega_J \approx 0.43$ and $\omega_p/\omega_0 \approx 0.17$, which
yielded a relatively large chromatic
dispersion (see Eq.\,(\ref{dispersion})) and thus a
sufficiently large phase mismatch for unwanted high-frequency
modes, $\omega_\pm$ and $\omega_{2p}$. As a result, the coherence lengths for
these waves, assuming $\omega_s \approx  \omega_i \approx 0.5\omega_p$,
were $\ell_+ =\ell_- = 54$ and $\ell_{2p} = 11$, respectively.
(By comparison, the pump wavelength was $\lambda_p = 2\pi/k_p
\approx 2\pi \omega_0/\omega_p = 38$.)
The phase mismatch for the basic 3WM process was found to be $\Delta k_0 \approx 0.014$,
which yielded a sufficiently large coherence length, $\ell_c =\pi/\Delta k_0
= 224 \gg \ell_\pm, \ell_{2p}$. Thus, the designed value of the poling period,
$m = 2 \ell_c$, amounted to 448.

The maximum gain of 20~dB, which was achieved
for QPM profile of type (ii)
at $\omega_s \sim 0.5\omega_p$, only roughly corresponds to the theoretical
value for QPM gain, whose formula is valid for $N/m \gg 1$.
Taking into account the effective value of the nonlinear coefficient,
$\beta_{\textrm{eff}} = 2 |\beta|/\pi \approx 0.16$, the exponential gain
factor is\:\cite{Zorin2016}
\begin{equation}
g = 0.25 \eta_{\textrm{eff}} \phi_p \omega_p/\omega_0 \approx 3.2\times 10^{-3},
\label{gain-theory}
\end{equation}
which yields a signal gain $G = 10 \log (\cosh^2 gN) \approx 22$\,dB $> 20$\,dB.
This overestimate can be explained by the fact that the number of poling periods
accommodated on the full length of our transmission line is not sufficiently
large, $N/m \approx 2$. Thus, obvious undulations of the growing signal that were
clearly seen in our simulations (see such plots, e.g., in Fig.\,2.4.2
in Ref.\,\onlinecite{Boyd2008}),
are significant on this scale.
However, little increase in the signal gain in such a short transmission line
is still possible at a small quasi-phase mismatch, $m \lesssim 2\ell_c$ (see dashed red
curve in Fig.\,4(b) calculated for a tapered
meander with a period of $m = 420$).

Comparison of the gains obtained for different
QPM profiles (see Fig.\,4(b)) shows that tapering of the profile
leads to an appreciable reduction in rippling (see curves (ii) and (iii)).
Although the sinusoidal profile (iii)  ensures the smoothest possible frequency dependence,
the gain level at a similar length $N$ and the similar pump amplitude $\phi_p$
is somewhat smaller (ca. $-2$~dB) than
for the tapered meander (ii) having larger $\eta_{\textrm{eff}}$.
Thus, exploiting profile (ii) seems to be a good trade-off between
maximum gain and smoothness of spectrum.
The resulting 3~dB bandwidth is ca. $0.4 \omega_p$ (cf. with the figure
for ideal 3WM amplifier of $0.47 \omega_p$ \cite{Tien1958,Cullen1960}).
For the given circuit parameters, the signal bandwidth is about 6~GHz.

In conclusion, we applied the optical QPM concept to a parametric amplifier for
traveling microwaves, designed a poled superconducting transmission line,
and demonstrated its efficiency by modeling.
We showed that poling in this JTWPA with 3WM can be realized using a
simple architecture with periodically inverted groups of identical rf-SQUIDs
in a uniform magnetic field.
Modeling using CMEs showed that this JTWPA can have a relatively large gain
over its wide bandwidth.
Moreover, the QPM approach can also be applied to JTWPAs
with 3WM based on SNAILs (Fig.~1(c)).
Finally, periodic inverting of every second rf-SQUID (or SNAIL) creates a simple design
of the transmission line with a fully suppressed nonlinearity of the $\chi^{(2)}$-type
($\beta = 0$) and widely tunable
Kerr coefficient  $\gamma$. \cite{Bell-Samolov2015,WenyuanZhang2017}
Such JTWPAs can be operated in pure 4WM regime with a tunable negative value of
$\gamma$, which was exploited in a recent study of a JTWPA
with reversed Kerr nonlinearity. \cite{Ranadive2021}

In summary, we believe, that by using the QPM concept in JTWPAs with 3WM,
these amplifiers will become the practical devices and support their wide
application in quantum technologies.

The author would like to acknowledge useful discussions with T. Dixon,
R. Dolata, and C. Ki{\ss}ling.
This work was partially funded by the Joint Research Project ParaWave of the
European Metrology Programme for Innovation and Research (EMPIR).
This project 17FUN10 ParaWave has received funding from the EMPIR programme co-financed
by the Participating States and from the European Union Horizon 2020
research and innovation programme.

The data that support the findings of this study are openly available
in Zenodo at https://doi.org/10.5281/zenodo.4732453, Ref. 38.

\end{document}